\documentclass[twocolumn]{article}
\usepackage[a4paper]{geometry}
\usepackage[utf8]{inputenc}
\usepackage{authblk}
\usepackage[numbers]{natbib}
\usepackage{graphicx}
\usepackage{xcolor}  
\usepackage{anysize}
\usepackage{wrapfig} 
\usepackage{caption}
\usepackage{amssymb}
\usepackage{float}
\usepackage{abstract}

\marginsize{2.5cm}{2.5cm}{2.5cm}{2.5cm}

\usepackage[compact]{titlesec}
\titlespacing{\section}{0pt}{2ex}{1ex}
\titlespacing{\subsection}{0pt}{1ex}{0ex}
\titlespacing{\subsubsection}{0pt}{0.5ex}{0ex}

\def\cmd{[cm$^{-3}$]}

\def\feii{Fe {\sc ii}}
\def\feiii{Fe {\sc iii}}
\def\hb{H$\beta$}
\def\mbh{$M\mathrm{_{BH}}$}

\def\civ{{C \sc{iv}} $\lambda$1549\/}
\def\aliii{Al {\sc iii} $\lambda$1860}
\def\siiii{Si {\sc iii]} $\lambda$1892}
\def\ciii{C {\sc iii]} $\lambda$1909}

\def\siiv{S {\sc iv} $\lambda$1397}

\def\ergs{[ers s$^{-1}$]}

\def\mgii{{Mg \sc{ii}} $\lambda$2800\/}
\def\oiii{{\sc{[O iii]}} $\lambda\lambda$4959, 5007\/}

\def\lledd{L/L$\mathrm{_{Edd}}$}
\def\nh{n$\mathrm{_{H}}$}

\def\Rblr{$R\mathrm{_\mathrm{BLR}}$}

\def\hubble{[km s$^{-1}$ Mpc$^{-1}$]}

\author[1*]{B. Czerny}
\author[1]{M.L. Mart\'inez-Aldama}
\author[2]{G. Wojtkowska}
\author[1]{M. Zaja\v{c}ek}
\author[3]{P. Marziani}
\author[4]{D. Dultzin}
\author[1]{M. H. Naddaf}
\author[1]{S. Panda}
\author[1]{R. Prince}
\author[5]{R. Przyluski}
\author[6]{M. Ralowski}
\author[1]{M. \'Sniegowska}
\affil[1]{Center for Theoretical Physics, Polish Academy of Sciences, Al. Lotnik\'ow 32/46, 02-668 Warsaw, Poland}
\affil[2]{Warsaw University Observatory, Al. Ujazdowskie 4, 00-478 Warszawa, Poland}
\affil[3]{INAF, Osservatorio Astronomico di Padova, Italy}
\affil[4]{Instituto de Astronom\'ia, UNAM, Mexico}
\affil[5]{Space Research Centre, Polish Academy of Sciences, Bartycka 18a, 00-716 Warsaw, Poland}
\affil[6]{Astronomical Observatory of the Jagiellonian University, Orla 171, 30–001 Krakow, Poland}
\affil[*]{Corresponding author: B. Czerny, bcz@cft.edu.pl}

\vspace{-20cm}

\title{Dark energy constraints from quasar observations}
\date{Key words: cosmology, dark energy, quasars}

\begin{document}

\twocolumn[
  \maketitle
  \begin{onecolabstract}
Recent measurements of the parameters of the Concordance Cosmology Model ($\Lambda$CDM) done in the low-redshift Universe with Supernovae Ia/Cepheids, and in the distant Universe done with Cosmic Microwave Background (CMB) imply different values for the Hubble constant (67.4 $\pm$ 0.5 \hubble\ from Planck vs 74.03 $\pm$ 1.42 \hubble\, Riess et al. 2019). This Hubble constant tension implies that either the systematic errors are underestimated, or the $\Lambda$CDM does not represent well the observed expansion of the Universe. Since quasars - active galactic nuclei - can be observed in the nearby Universe up to redshift z $\sim$ 7.5, they are suitable to estimate the cosmological properties in a large redshift range. Our group develops two methods based on the observations of quasars in the late Universe up to redshift z$\sim $4.5, with the objective to determine the expansion rate of the Universe. These methods do not yet provide an independent measurement of the Hubble constant since they do not have firm absolute calibration but they allow to test the $\Lambda$CDM model, and so far no departures from this model were found.  \end{onecolabstract}
]


\section{Introduction}
The cosmological parameters can be estimated from different sets of data at various redshifts, but if the standard $\Lambda$CDM  (Lambda-Cold Dark Matter) model is valid, they can always be represented by the current (zero redshift) values. The final results from the Planck mission, based on the analysis of the Cosmic Microwave Background (CMB) do not indicate any tension with the standard model, and give the value of the $\Omega_m = 0.315 \pm 0.007$ \citep{Planck2020} and the Hubble constant $H_0 = 67.4 \pm 0.5$ \hubble. Many of the measurements done in the local Universe ($z < 10$) are in significant disagreement with these $\Omega_m$ or $H_0$ values (e.g. \cite{riess_review_2019}) while other measurements, also local, are still roughly in agreement with the results from Planck (e.g. the last results from gravitational waves \cite{LIGO2019}).

Therefore various probes and methods are needed to confirm, or to reject, the hypothesis that the $\Lambda$CDM model does not well describe the Universe, and the evolving dark energy is needed instead of the cosmological constant. Quasars (QSO) are very attractive cosmological probes, since they cover a broad range of redshifts, from nearby sources (referred to as Active Galactic Nuclei, AGN) to most distant objects at redshift above 7  \citep{mortlock2011,banados2018}. They also do not show significant evolution with redshift \citep{onoue2020}.  

\section{Two methods for using quasars in cosmology}

We are currently using two methods of turning quasars into {\it standardizable} candles. The first method is based on the radius-luminosity relation and the second method is based on super-Eddington sources.

\subsection{Method based on radius-luminosity relation for the BLR}

The reverberation mapping technique is based on the long-term monitoring of a source in order to determine the time response ($\tau_{\rm BLR}$) of the emission line to the continuum variations \citep{peterson2004}. The most important result from the reverberation mapping studies is the correlation between the continuum luminosity ($L$)  and the distance (\Rblr) where the emission line is emitted in the broad line region (BLR). This relation is known as the Radius-Luminosity relation (RL) and it is given approximately by \Rblr$\propto L^{0.5}$. The reverberation mapping studies require extensive use of telescope time to achieve high quality results, thus only $\sim120$ sources have been analyzed with this technique until date. Most of the monitoring are based on the optical \hb\ for low-redshift sources, while for high redshift regimes, due to the Doppler shift, the monitoring are focused on the UV emission lines such as \mgii, \civ\ and \ciii. 

For many years the RL relation showed a low scatter ($\sigma_{\rm rms}\sim0.13$ dex) \citep{kilerci2015}, which ensured its use in the determination of the black hole mass (\mbh). However, the inclusion of new sources, particularly those radiating close to the Eddington limit  (high accretion rates), has led to a much larger scatter, clearly related with the accretion rate \citep{du2015, du2018}. Some corrections based on the accretion rate \citep{martinez-aldama2019} and independent parameters such as the \feii\ strength or the amplitude of variability \citep{du2019, dallabonta2020, martinez-aldama2020} (in turn correlated with the accretion rate) have been proposed to correct this effect,
allowing to reduce the scatter.

\begin{figure}[H]
\includegraphics[width=0.5\textwidth]{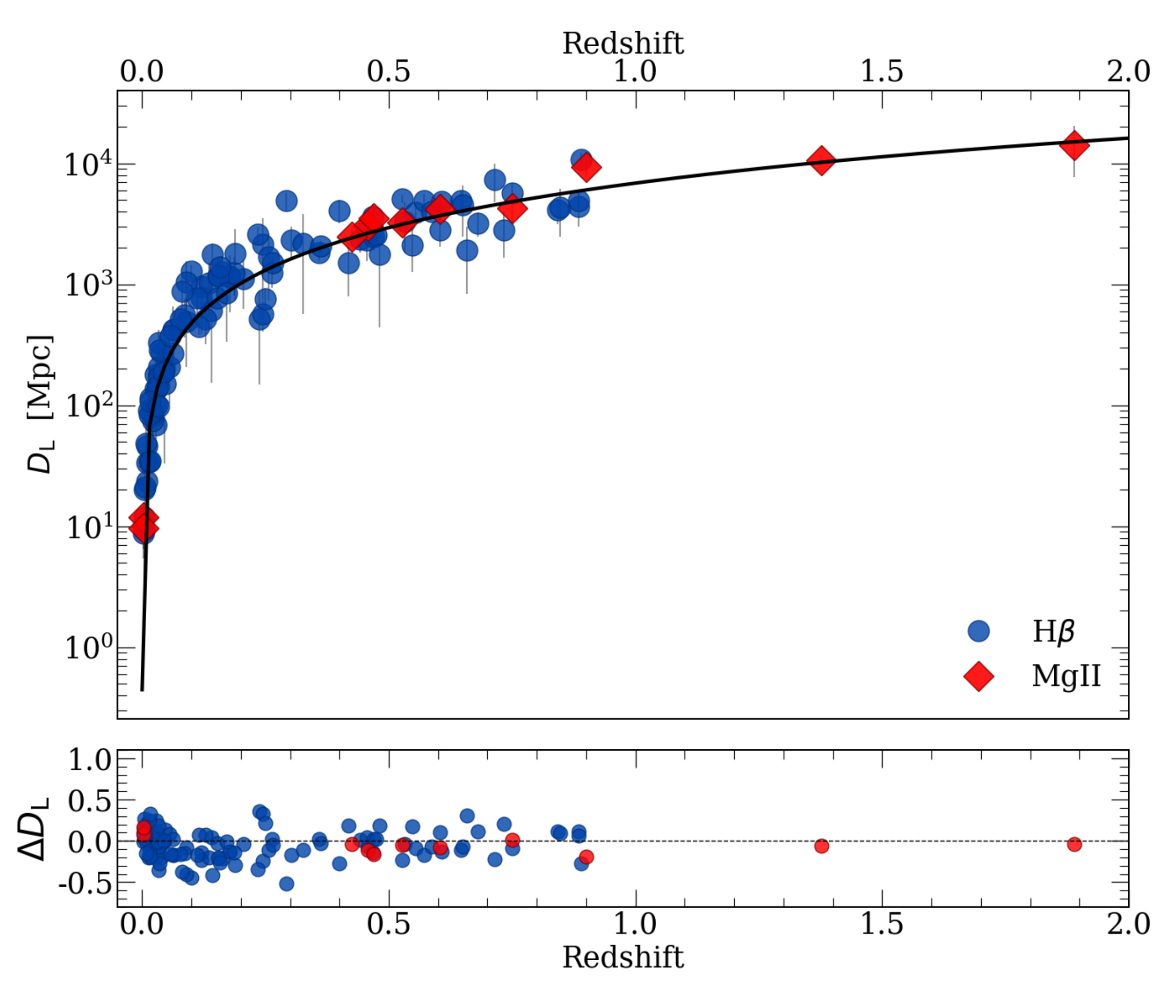}\\
\includegraphics[width=0.5\textwidth]{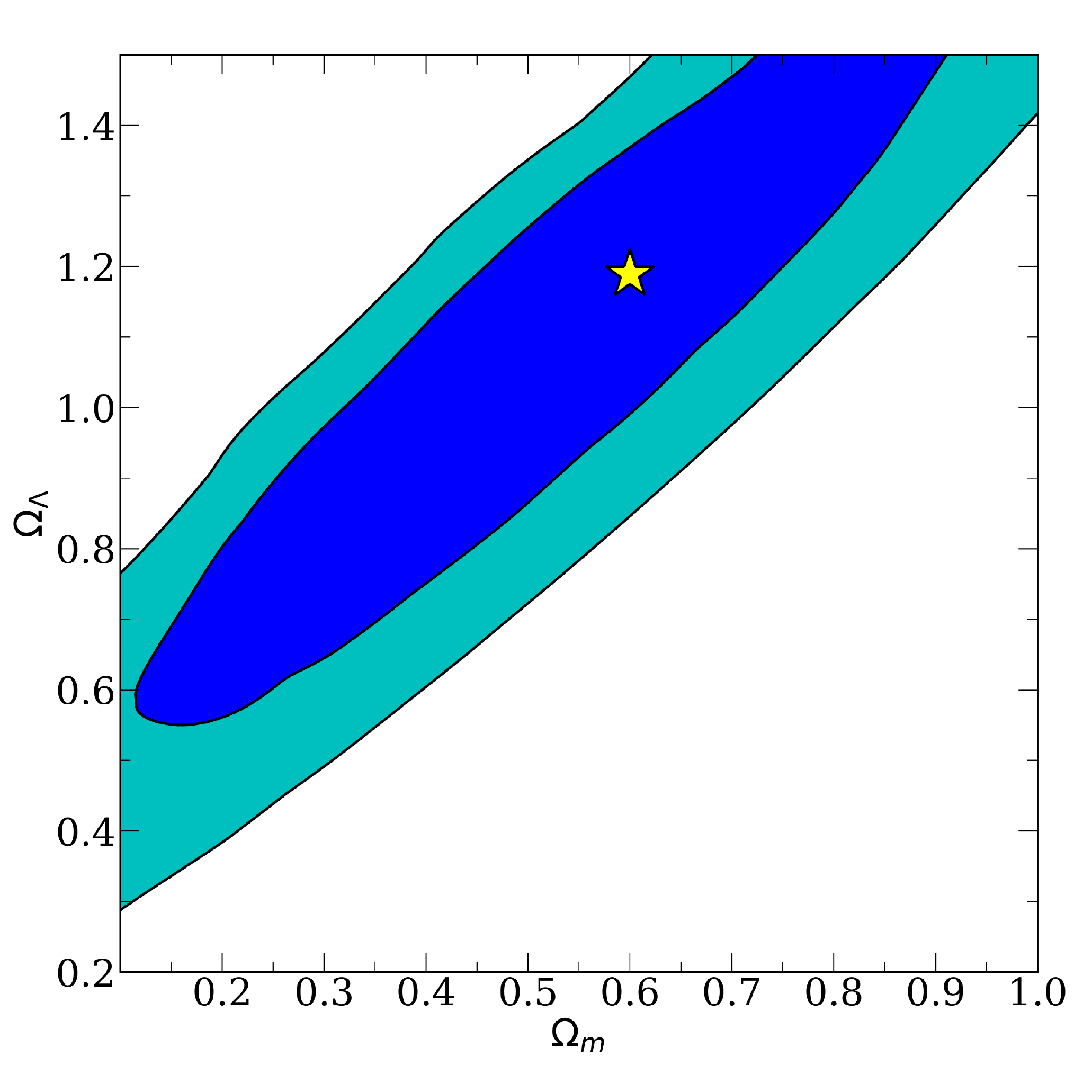}
\caption{Top panel: Quasar Hubble diagram using the reverberation mapped sources. Blue circles and red diamonds correspond to the \hb\ and \mgii\ sources, respectively. Black line marks the $\Lambda$CDM model for flat cosmology, $\Omega_m = 0.297$, $H_0 = 67.5$ \hubble . Bottom subpanel shows the residuals. Bottom panel: Confidence contours at 68$\%$ (cyan) and 95$\%$ (blue) for $\Omega_m$ and $\Omega_\Lambda$ for general $\Lambda$CDM model based on $\chi^2$ fitting where the best $\Omega_m$ and $\Omega_\Lambda$ are represented by the yellow symbol.  }
\label{fig:hubble_RL}
\end{figure}

Besides, the Radius-Luminosity relation offers the possibility to determine the luminosity independently of the redshift \citep{watson2011, haas2011}, and to estimate the cosmological parameters. However, in \citep{martinez-aldama2019} the errors for $\Omega_{m}$ and $\Omega_{\Lambda}$ based on available  \hb\  were still very large, despite the corrections for the accretion rate. In the present paper, we make the following important modifications. First, we combine the previous sample with \mgii\ reverberation-mapped sources. The new sample includes two sources monitored by us with The Southern African Large Telescope (SALT) during 6 years \citep{czerny2019, zajacek2020}. Next,  since the errors of the time delay measurement are highly asymmetric, we use the $\chi^2$  statistics to determine the cosmological parameters, we use the method of \cite{barlow2004} instead of a simple symmetrization of the errors.

We also modified the approach to the R-L relation in case of the \hb\ sample which is extremely heterogeneous. We treated the coefficients of this relation as arbitrary, and minimized the total $\chi^2$ fit to a flat cosmological model, with Hubble constant fixed at 67.5 \hubble.  For  \mgii\ sample, we used the R-L parametrization given by Eq. 9 in \cite{zajacek2020} since this sample was analysed in a more uniform way. While fitting the flat cosmology model (both samples combined), we applied the sigma-clipping approach, and we removed the sources which showed a departure by more than 3 sigma from the best-fit. All the removed sources were from the \hb\ sample: Mrk 493, J074352.02+271239.5, Mkn 509, MCG+08-11-011, J142103, J142043, J141123, and J142052. Thus our total sample has now 120 objects. We then refitted the Hubble diagram. The best-fit returned the best R-L parametrization of \hb\ sample as $\log L_{5100} = 1.489 \log \tau_{\rm corr} - 2.222$, where $\tau_{\rm corr}$ is the time delay corrected by the accretion rate effect \citep{martinez-aldama2019}. For the flat cosmology, we obtained the best-fit value $\Omega_m = 0.297^{+0.060}_{-0.054}$ (see Fig.~\ref{fig:hubble_RL}, top panel). This value is fully consistent with the value $0.3153 \pm 0.0073$ from Planck \citep{Planck2020} for flat cosmology, and the error in our new result is much smaller than obtained by \cite{martinez-aldama2019}, although still much larger than from Planck. This illustrates that the method could be powerful, if the sample is more uniformly analysed from the very beginning. If we do not assume a flat cosmology, 2-D contour errors are still large (see Fig.~\ref{fig:hubble_RL}, bottom panel), although considerably smaller than in \cite{martinez-aldama2019}. The best-fit Planck values are well within the  1$\sigma$ error, so we do not see any tension with the results based on Cosmic Microwave Background.

\subsection{Method based on Super-Eddington sources}

Quasars radiating close to the Eddington limit are known as {xA-QSO} or super--Eddington sources \citep{marzianisulentic2014, wangSEAMBH2014}. This QSO population shows peculiar spectral and photometric properties, which differentiate them from the rest of the QSO population and make them easy to identify in catalogs like  SDSS or  the upcoming Vera Rubin Observatory's Legacy Survey of Space and Time (LSST). In the optical range, they are the strongest \feii\ emitters and do not show a strong contribution of narrow emission lines such as \oiii\ \citep{negrete2018}. xA-QSO show the strongest outflows in the high ionization lines mostly observed in the UV emission lines like \civ\ or \siiv\ \citep{sulentic2017}, although in the most extreme cases the strong radiation forces provoke the presence of  outflows in low-ionization lines such as \hb\ or \aliii. According to the photoionization models, their broad line regions show large densities (\nh$=10^{12-13}$ \cmd), low-ionization parameters (log $U< -2$) and high metallicities ($Z\sim10Z_\odot$) \citep{martinez-aldama2018, sniegowska2020, panda2020a, panda2020b}. In addition, Super-Eddington sources also show remarkably low optical variability and time delays shorter than the predicted by the RL relation \citep{du2016}. The UV flux ratios \aliii/\siiii$>$0.5 and \ciii/\siiii$<$1.0 have shown a high effectiveness as selection criteria to identify xA-QSO sources at high-redshift, while at-low redshift xA-QSO typically show \feii/\hb$>$1.0   \citep{marzianisulentic2014}.

In {xA-QSO}, despite the rise of the accretion rate, the luminosity saturates toward a limiting value, since the accretion efficiency decreases. Thus the ratio luminosity--black hole mass ($L$/\mbh) does not change and they can be considered as “Eddington standard candles”. A similarity in the physical conditions (density, ionization parameter, metallicity) of the BLR is expected because them belonging to the same population, therefore a generalization of all of them can be considered \citep{panda2019b}. Since the low-ionization lines are less affected by the strong radiation forces, emission lines like \hb\ and \aliii\ are excellent candidates for virial estimators. Based on these assumptions, it is possible to determine the luminosity distances independently of redshift and get an estimation of the matter and energy content in the Universe.

The previous xA-QSO sample included $\sim200$ objects at redshift $z<2.7$ \citep{dultzin2020}. In order to increase the redshift range, we considered the most recent edition of the The Sloan Digital Sky Survey Reverberation Mapping (SDSS-RM) catalog \citep{shen2019}. This catalog includes the automatic measurements of the most important UV emission lines for 549 sources with 44.1$<{\rm log }L_{1700}<$46.9 \ergs\  at $0.9<z<4.3$, where $\sim20\%$ show high accretion rates, so they can be considered as Super-Eddington candidates.  In the xA sources the \feiii$\lambda1914$ shows an important contribution, hence a good deblending of the \ciii\ and \feiii\ $\lambda1914$ is required. Unfortunately, the SDSS catalog does not include \feiii\ in their multicomponent fittings, so not all the sources satisfy the selection criteria to identify them as the xA. So for the first test, we select the sources based on the Eddington ratio (\lledd{}$>0.2$) estimated from the \aliii\ based black hole mass. After removal of some objects with extreme FWHM$_{\rm AlIII}$ values, our final sample includes 88 objects at $1<z<4.5$.

\begin{figure}[H]
\centering
\includegraphics[width=0.5\textwidth]{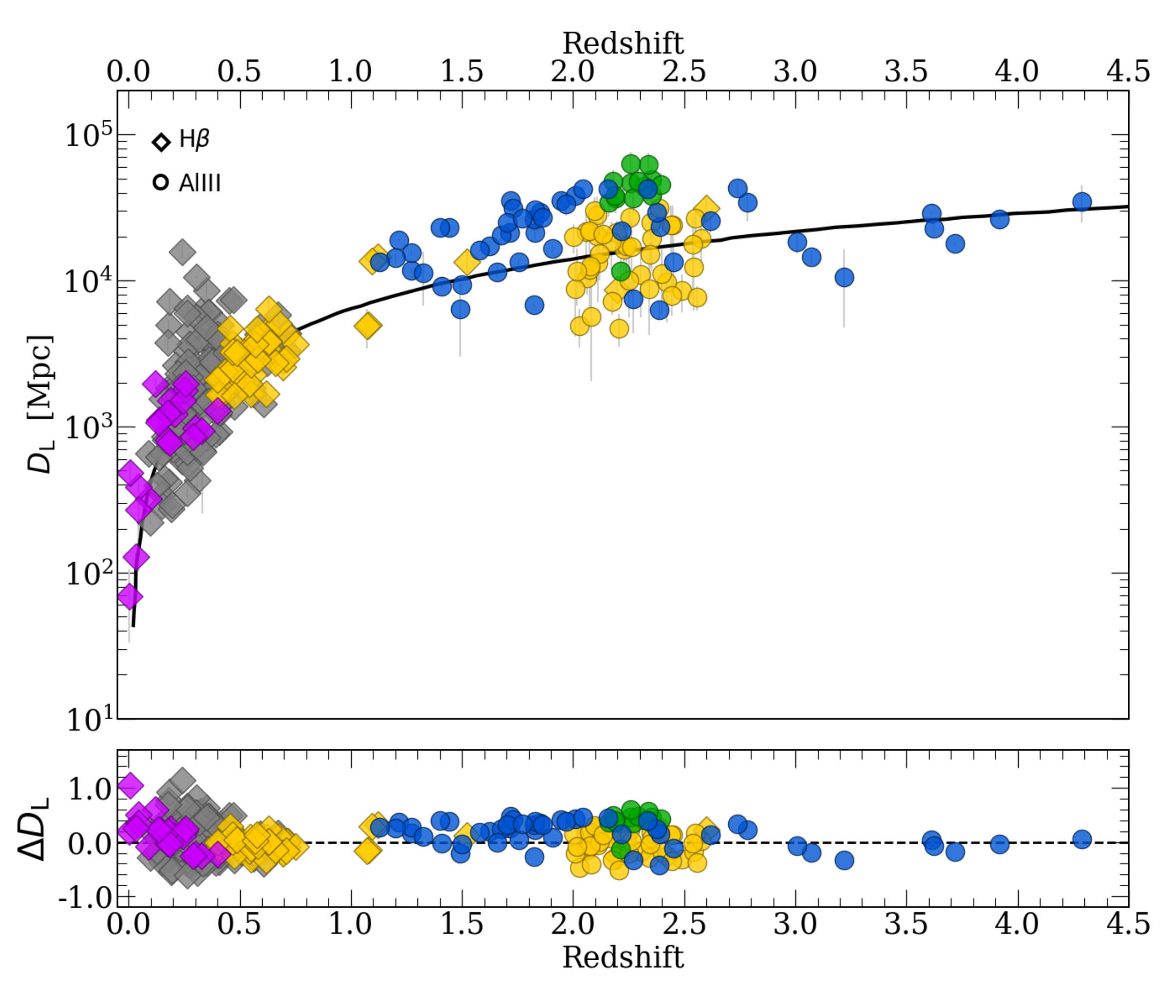}
\caption{Quasar Hubble diagram using the Super-Eddington sources. Diamonds correspond to the measurements from the \hb, while circles belong to UV \aliii\ emission line. Purple, gray, yellow, green and blue symbols correspond to the xA sources from the SEAMBH project \citep{du2019}, \citep{negrete2018}, \citep{marzianisulentic2014},  \citep{sniegowska2020} and SDSS-RM \citep{shen2019} samples, respectively. The best-fit line shows the flat model, with $H_0$ from Planck, and best fit $\Omega_m = 0.290$.}
\label{fig:hubble_xA}
\end{figure}

In order to determine the cosmological constant $\Omega_m$ and $\Omega_\Lambda$, we combine the previous xA samples such as the super-Eddington sources from the super-Eddington accreting
massive black holes (SEAMBHs) project with the most recent measurements from \citep{sniegowska2020}. The quasar Hubble diagram with the super-Eddington sources is shown in Fig.~\ref{fig:hubble_xA}. We adopt the scaling of the virial estimator to be consistent with Planck $H_0$ value, and we assume the flat cosmology.
In this case we obtain $\Omega_m = 0.290^{+0.048}_{-0.043}$, fully consistent with the Planck results despite the fact that quasars cover the redshift range from nearby sources to almost 4.5.



\section{Discussion}

Here we presented the most recent results based on the two methods for applying quasars to constrain the expansion rate of the Universe. Our methods, as for now, are not based on absolute scaling so they cannot predict the value of $H_0$. In principle, such an absolute scaling can be achieved. For method (i) it  would require an independent measurement of the dust temperature at the BLR onset, and the development of a 3-D BLR model, which is in progress (see e.g. \cite{naddaf2020}). For method (ii), we would need an absolute scaling of the radius-luminosity relation, also establishing the mean density of the BLR (see e.g. \cite{marzianisulentic2014, panda2019b}). At this stage, we fixed the value of the Hubble constant at the Planck value and tested, whether the redshift dependence of the luminosity distance is consistent with the standard $\Lambda$CDM model, and whether the remaining cosmological parameters derived from quasar data are consistent with Planck values.

So far, within the available accuracy, our values of $\Omega_m$ are fully consistent with the Planck value for the flat Universe despite the fact that second method extends up to the redshift 4.5. Thus we do not support the claim of the tension with the standard model based on Supernovae Ia with absolute scaling in turn predominantly based on Cepheid stars \cite{riess2018}. Our results from method (i) are consistent with the tension found by \cite{lusso2020} since they claim to see departures only above the redshift 1.5 - 2, and method (i) does not go this far. As for the method (ii), we have many sources up to redshift 2.5, but indeed very few above 2.5, and our method of analysis was not yet optimized by an outlier removal through sigma-clipping. Further studies are clearly needed for this method, both with the current data and eventually by increasing the number of high redshift quasars.


\section*{Acknowledgement}
The  project  is  partially based  on  observations made with the SALT under programs 2012-2-POL-003, 2013-1-POL-RSA-002,  2013-2-POL-RSA-001,  2014-1-POL-RSA-001,  2014-2-SCI-004,  2015-1-SCI-006,  2015-2-SCI-017, 2016-1-SCI-011, 2016-2-SCI-024, 2017-1-SCI-009, 2017-2-SCI-033, 2018-1-MLT-004 (PI: B. Czerny). 
The authors acknowledge the financial support by the National Science Centre, Poland, grant No.~2017/26/A/ST9/00756 (Maestro 9), and by the Ministry of Science and Higher Education (MNiSW) grant DIR/WK/2018/12. The Polish participation in SALT is funded by grant No. MNiSW DIR/WK/2016/07.

\bibliographystyle{unsrtDOI}
\bibliography{references}
\end{document}